\documentclass[aps,prb,twocolumn,superscriptaddress,floatfix,amsmath,amssymb]{revtex4}
\usepackage{epsfig}
\usepackage{amssymb}

\makeatletter
\def\@dotsep{1024} 

\def\contentsline#1#2#3#4%
{\csname l@#1\endcsname{#2}{}}
\def\jcp@caption#1{\revtex@caption[#1]{}} 
\makeatother

\begin{document}

\title{Comparison of structure and transport properties of concentrated hard and soft sphere fluids}
\author{Erik Lange}
\affiliation{University of Konstanz, D-78457 Konstanz, Germany}
\author{Jose B. Caballero}
\affiliation{Group of Complex Fluids Physics, Department of Applied Physics, University of Almeria, 04120 Almeria, Spain}
\author{Antonio M. Puertas}
\affiliation{Group of Complex Fluids Physics, Department of Applied Physics, University of Almeria, 04120 Almeria, Spain}
\author{Matthias Fuchs}
\affiliation{University of Konstanz, D-78457 Konstanz, Germany}
\date{\today}

\begin{abstract}

Using Newtonian and Brownian dynamics simulations, the structural and transport properties of hard and soft spheres have been studied. The soft spheres were modeled using inverse power potentials ($V\sim r^{-n}$, with $1/n$ the potential softness). Although, at constant density, the pressure, diffusion coefficient and viscosity depend on the particle softness up to extremely high values of $n$, we show that scaling the density with the freezing point for every system effectively collapses these parameters for $n\geq 18$ (including hard spheres), for large densities. At the freezing points, the long range structure of all systems is identical, when length is measured in units of the interparticle distance, but differences appear at short distances (due to the different shape of the interaction potential). This translates into differences at short times in the velocity and stress autocorrelation functions, although they concur to give the same value of the corresponding transport coefficient (for the same density to freezing ratio); the microscopic dynamics also affects the short time behaviour of the correlation functions and absolute values of the transport coefficients, but the same scaling with the freezing density works for Newtonian or Brownian dynamics. For hard spheres, the short time behaviour of the stress autocorrelation function has been studied in detail, confirming  quantitatively the theoretical forms derived for it.

\end{abstract}

\maketitle

\newpage

\section{Introduction}

Monodisperse spherical particles with short range repulsions show a simple phase diagram, with only one fluid at low density and one crystal phase at high density, and a first order transition in between \cite{Reiss86}. The paradigmatic case is the system of hard spheres (HS), where the interaction potential is infinite whenever particles overlap and zero elsewhere, and therefore there is no energy scale in it. In HS, the crystallization transition was first recognized by Alder and Wainwright \cite{Alder57}, and has the freezing and melting points at volume fractions $\phi_f=0.494$ and $\phi_m=0.545$, respectively \cite{Rintoul96}. Additionally, a good approximation for the equation of state for HS fluid was given by Carnahan and Starling using the virial expansion \cite{Carnahan69}, tested experimentally using sedimentation of screened charged colloids \cite{Piazza93}. For slightly soft spheres (SS), where interpenetration is not yet an issue, the phase diagram changes quantitatively, shifting the fluid-solid coexistence, and introducing temperature as a new variable (the energy scale, $\epsilon$, is set by the potential, and $\epsilon/k_BT$, where $k_BT$ is the thermal energy, is coupled to the volume fraction). Inverse-power potentials ($V(r)\sim r^{-n}$) have been widely used to model SS, and the phase diagram studied as a function of the softness by simulations \cite{Agrawal95b,Agrawal95} and theories
\cite{Rogers84,Carbajal08}. Because of the qualitative similarity of the phase diagram of SS with that of HS, a mapping of the former to the latter has also been tried by means of effective diameters \cite{Andersen71,hansen86,Heyes97}. However, an interesting pathology arises in HS which is absent in SS: the elastic modulus for large frequencies diverges in HS \cite{Heyes94,Lionberger94,Verberg97}.

Experimentally, different colloidal systems have been used to model HS, such as silica \cite{Werff89} or latex particles \cite{Piazza93,Pusey86}, or collapsed microgel \cite{Kasper98} or core-shell\cite{Crassous08} particles. Because the hard sphere interaction is an idealization of quasi-rigid spheres, these systems are usually referred to as "nearly hard spheres", and the question of their softness arises naturally \cite{Bryant02}. In most cases, additionally, a (generally thin) polymer layer is adsorbed onto the particles to provide steric stability, what increases the softness of the particles \cite{Bryant02}. Even more, experiments with really soft particles, such as swollen microgels \cite{Senff99} or star polymers \cite{Watzlawek99}, can show new phases due to interpenetration, but also phase diagrams qualitatively similar to the HS one for the harder-particle limit. However, the short range interaction is not directly accessible experimentally \cite{Bryant02}, and exact knowledge of the true softness of the particle is missing. It is therefore important to know what properties depend on the softness of the particle, and if there is a simple scaling for different systems with different softness.

Recent simulations of inverse power potentials have shown that the transport coefficients depend strongly on the particle softness, and the HS limit is reached for large values of $n$ ($n>72$) \cite{Heyes05}. It has been shown that the time-correlation functions (and transport coefficients) can be approximately scaled using the exponent $n$ \cite{Powles00,Dufty02}, but neither the scaling is perfect nor the exponent is known experimentally. On the other hand, the transport coefficients behave quasi-linearly with the inverse packing fraction for different softness, what can indicate that the key parameter for mapping the systems is the free volume, but a detailed analysis of the parameters shows inconsistencies (for instance, the free volume vanishes at different density for every transport coefficient) \cite{Heyes05}. It is desirable to find a simple mapping from soft particles to HS, if it exists, that can be applied in experimental systems.

The freezing point, on the other hand, is well defined fundamentally, and can be identified with several well-known, albeit approximate, criteria \cite{Rosenfeld81}, independently of the interaction potential, such as the Hansen-Verlet criterion, which uses the height of the neighbour peak in the structure factor \cite{Hansen69}, and a dynamical one, proposed by L\"owen and co-workers, based on the decrease of the long time diffusion coefficient \cite{Lowen93}. In computer simulations, the freezing point can be determined accurately using Gibbs-Duhem integration from HS \cite{Agrawal95}. Therefore, in this work, we propose to use the density at freezing to scale the results. Using computer simulations, we show that the structural and dynamical quantities of interest of inverse-power potentials can be rationalized when plotted against the scaled density, $\rho/\rho_{\text{freezing}}$, for large density at identical temperature. The long range structure is almost identical for systems with equal $\rho/\rho_{\text{freezing}}$, but differences appear at short distances due to the different interaction potentials. The pressure, which shows a rather complicated trend when studied at constant density, increases monotonously with the softness at constant $\rho/\rho_{\text{freezing}}$. The diffusion coefficient and viscosity can be collapsed for $n \geq 18$, including HS, and the time correlation functions only show differences at short times, giving different shear moduli at large frequencies. Our conclusion is therefore that the relative density to the freezing transition is the key parameter governing the structure and dynamics of the system, for large enough $n$ and density. This poses a simple criterion that can be useful for experiments.

\section{Simulation details}

We make simulations of different monodisperse systems: hard spheres and soft spheres with different "softness". Whereas the interaction potential is continuous for soft spheres, it is not for hard spheres, what makes an important difference in the computational method. The microscopic dynamics is Newtonian, but we have also performed some Brownian dynamics or damped Newtonian dynamics to make more direct contact with experimental colloids.

\subsection{Hard spheres}
Hard spheres only interact upon contact, when they collide elastically due to their excluded volume.
Given such a discontinuous potential we resort to an event-driven algorithm
for the hard-sphere simulations. The method relies on the fact that in between collisions
particles move undisturbed and in a straight line. It is therefore advantageous to only process
the collisions when they occur. This is the basic feature of an event-driven algorithm: it does
not evolve with a fixed time step, but instead time propagation is determined by the events in
the simulation --- the hard sphere collisions in the case of Newtonian dynamics.
In this paper we have modeled the interactions of the hard spheres by elastic collisions.

The simulation consists of 1000 particles in a cube to which we apply standard periodic boundary
conditions. Simulations are started from a simple cubic lattice. Velocities are then drawn at random from a Gaussian
distribution to let the system evolve towards equilibrium. From such an equilibrated
configuration all further simulations are run.

The standard deviation of the velocity determines the temperature of hard spheres:
$\left\langle v^{2}\right\rangle=\frac{3}{m}k_{B}T$.
Inherently Newtonian dynamics does not show any dependence on temperature.
Changing the temperature leads to a different timescale, but does not modify the dynamics qualitatively.
It is therefor the density alone that determines the physical behaviour of the hard spheres.

In all our simulations the diameter $\sigma$ determines the length scale and the particle mass $m$ the
unit of mass, while $k_{B}T$ may vary. With these conventions the time is given in units of
$\tau=t/\left(\sigma\sqrt{\frac{m}{k_{B}T}}\right)$ wherever it appears.

We have introduced Brownian motion by incorporating a random thermostat. This
feature introduces a new timescale $\tau_B$ on which the random kicks of
the solvent are mimicked by drawing new velocities for each particle from a Gaussian distribution\cite{scala07}.
This timescale must be chosen such that it effectively generates diffusion between collisions.
Given densities up to the freezing point we find that $\tau_B=0.017$ is sufficiently small (smaller values of $\tau_B$ cause too slow microscopic dynamics, increasing the simulation time \cite{Lange09}).

\subsection{Soft spheres}

\begin{figure}
\psfig{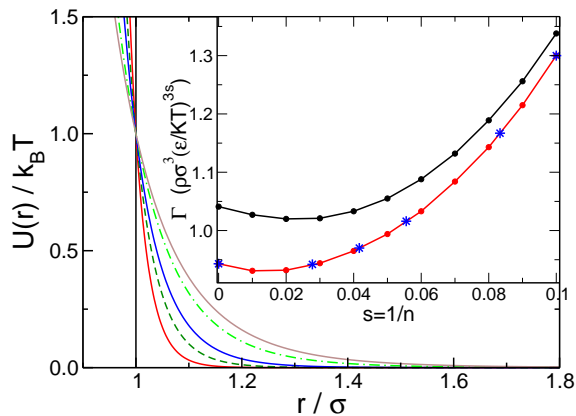}
\caption {Pair interaction potentials used in this work, with decreasing $n$ (increasing softness) from below: $n=36$, $24$, $18$, $12$ and $10$, and hard sphere. Note the long range of the repulsion for the lowest $n$. The inset shows the freezing and melting lines in terms of $\Gamma$ \cite{Agrawal95}. The stars indicate state points checked in simulations.}
\label{potential}
\end{figure}

For soft spheres we use a continuous repulsive potential:

\begin{equation}
U_{sc}(r)\:=\:\epsilon \left(\frac{r}{\sigma}\right)^{-n}
\end{equation}

\noindent where $\epsilon$ sets the energy scale and $s=1/n$ is the {\sl softness} of the potential. In this work, we change $n$ from $n=10$ to $n=36$, what provides a wide enough range to show clear effects from the softness of the potential, and the crossover to {\sl effective} hard spheres. Figure \ref{potential} presents the interaction potentials for $n=36$, $24$, $18$, $12$ and $10$ and the hard core repulsion. Note that as $n$ decreases (the softness increases), the potential has a longer range, and for $n=10$, the interaction energy is negligible only for distances above $r \approx 2\sigma$.

Since the interaction potential is soft, the particle volume cannot be unambiguously defined, and thus the volume fraction is not a convenient control parameter for these systems, although it is commonly used in experiments. Instead, states in potentials with different softness can be compared using the {\sl effective density} defined by the parameter $\Gamma=\rho \sigma^3 (\epsilon/k_BT)^{3/n}$, where $\rho$ is the number density. In this work, the temperature is fixed to $k_BT=\epsilon=1$, and variations in $\Gamma$ are caused by changes in the number density exclusively.

Conventional molecular dynamics are performed integrating the Newton's equations of motion with velocity rescaling to maintain constant temperature. The time step for the integration is $\delta t=0.002$, in standard units of $\sigma \sqrt{m/\epsilon}$. All potentials have been truncated for $V<10^{-3}$.

Damped Newtonian dynamics have been performed also, introducing friction and fluctuation (Brownian) forces. For particle $j$, the equation of motion is now Langevin equation:

\begin{equation}
m\ddot{\vec{r}}_j\:=\:\sum_i \vec{F}_{ij} - \gamma \dot{\vec{r}}_j + \vec{f}_j
\end{equation}

\noindent where $\vec{F}_{ij}$ is the interaction force between particles $i$ and $j$, $\gamma$ is the friction coefficient with the solvent, and $\vec{f}_j$ is the Brownian force. The fluctuating and viscous terms are related by the fluctuation dissipation theorem: $\langle \vec{f}_j(t') \vec{f}_i(t)\rangle=6 k_BT \gamma \delta_{ij} \delta(t-t')$. In this work, $\gamma=20 \sqrt{\epsilon m}/{\sigma}$ which gives a strong damping (the momentum relaxation time scale, $m/\gamma \sim 0.05$, is slightly smaller than the typical collision time between particles). The equations of motions are integrated using the algorithm developed by Heun \cite{paul95}, with a time step $\delta t=0.0005$.

\section{Results and Discussion}

Agrawal and Kofke used Gibbs-Duhem integration to compute the solid-fluid phase boundary for SS modeled with inverse-power potentials, starting from the FCC-fluid coexistence in HS \cite{Agrawal95}. Their results for the freezing transition are presented in Table I (and in the inset to Fig. \ref{potential}) for different values of $n$ and the HS system. Note that freezing density does not evolve monotonously with the potential softness (at constant temperature $k_BT=\epsilon$, $\Gamma=\rho \sigma^3$): upon decreasing $n$ from infinite (increasing softness starting from HS), $\Gamma_{\text{freezing}}$ first decreases, and then increases for $n<36$.

\begin{table}
\begin{center}
\begin{tabular}{cllcllc} \hline \hline
$n$ & & & $s=1/n$ & & & $\Gamma_{\text{freezing}}$ \\ \hline \hline
$\infty$ & & & 0 & & & 0.943 \\
36 & & & 0.0277778 & & & 0.942 \\
24 & & & 0.0416667 & & & 0.970 \\
18 & & & 0.0555556 & & & 1.016 \\
12 & & & 0.0833333 & & & 1.167 \\
10 & & & 0.10 & & & 1.300 \\\hline \hline
\end{tabular}
\caption {Tabulated data of the equilibrium phase diagram from Agrawal and Kofke \cite{Agrawal95}.}
\end{center}
\end{table}

\begin{figure}
\psfig{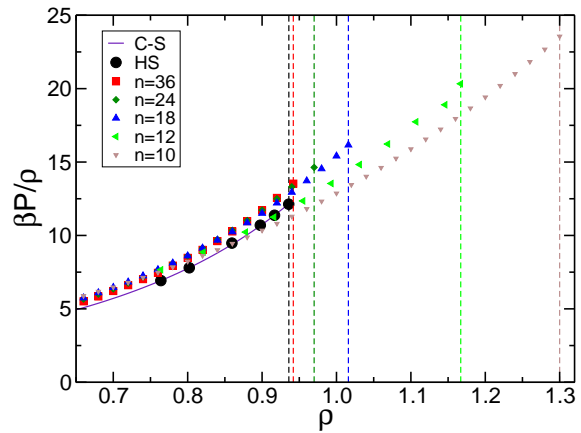}
\caption {Pressure of the different systems studied here, as labeled, and Carnahan-Starling equation (magenta line). The vertical dashed lines indicate the freezing densities from Agrawal and Kofke \cite{Agrawal95}.}
\label{pressure}
\end{figure}

Figure \ref{pressure} presents the pressure in the fluid phase up to the freezing point for different values of $n$. Whereas at low density the pressure is almost independent of the softness, important differences are noticed at high density, close to freezing. In particular, every system has a different pressure at fixed density, or at its own freezing point. Also, at low density (not shown in the figure) the pressure increases weakly with the softness, whereas at high ones it decreases.

\begin{figure}
\psfig{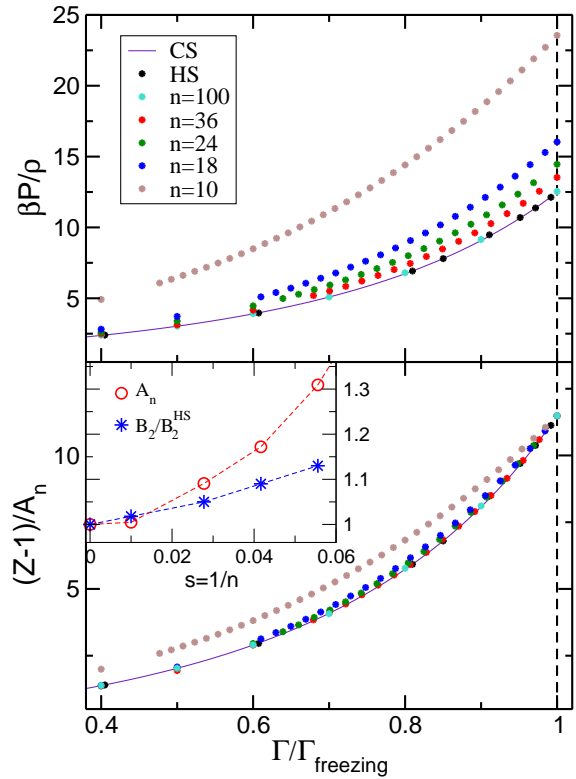}
\caption {Upper panel: Pressure as a function of the scaled density for different potential softness, as labeled, and Carnahan-Starling equation (magenta line). Lower panel: Scaled pressure vs. scaled density for the same values of $n$. The inset shows the scaling factor for the pressure (red circles) and the ratio of second virial coefficients to the hard sphere system (blue asterisks) for different $n$.}
\label{pressure-scaling}
\end{figure}

Contrary to other works, we will not try to rationalise the complex picture presented in Fig. \ref{pressure} with the so-called {\sl effective diameter}
\cite{Andersen71,Heyes97}, mapping the pressure to the hard-sphere one, or other theoretical approximations \cite{Rogers84}, but we will show that a convincing scaling of the data can be performed using the freezing density, showing that the hard sphere limit can be reached, within this scaling, for $n$ as low as $n\approx 24$. In the upper panel of Fig. \ref{pressure-scaling} we present the pressure as a function of the scaled $\Gamma$. A monotonous trend of the pressure with the potential softness is now evident at all shown densities.

In the lower panel the pressure has been scaled with the value at freezing, showing almost perfect collapse of the data for $n \geq 18$. The inset presents the scaling factor, $A_n$, (ratio of the pressure at freezing for SS to pressure for HS) and the ratio of the second virial coefficient, $B_2$, to the hard sphere value, $B_2^{HS}$, as a function of the potential softness. These two quantities disagree, showing that a mapping based on the second virial coefficient would not find such a good scaling, concluding that soft spheres behave as hard ones only for very large values of $n$. The a priori surprising finding emerges, that a scaling of soft spheres onto hard ones works better at high densities, where the closeness to freezing provides a common scaling variable, which collapses the data much better than the second virial coefficient would.

\begin{figure}
\psfig{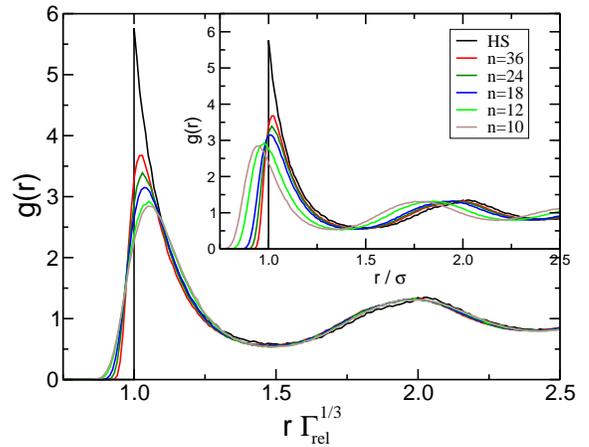}
\caption {Pair distribution function of the soft and hard spheres at freezing (see table I), for different softness: from top to bottom in the nearest neighbour peak $HS$, $n=36$, $n=24$, $n=18$, $n=12$ and $n=10$. The inset shows the raw distribution as a function of the average distance measured in units of $\sigma$, and the main panel in units of $\rho^{-1/3}$ (the hard sphere system has been left unchanged). }
\label{gr}
\end{figure}

\begin{figure}
\psfig{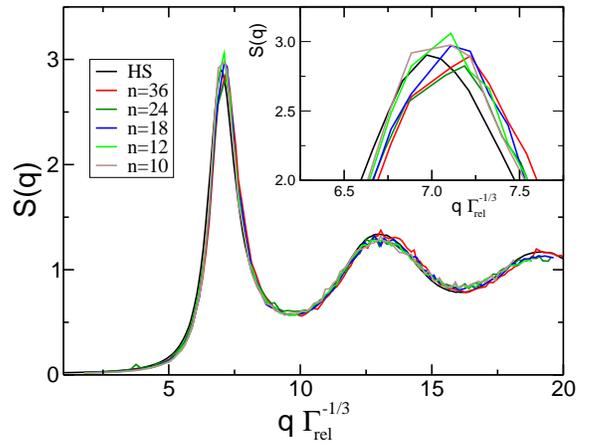}
\caption {Structure factor of the soft and hard spheres at freezing (see table I), with the same colour code as Fig. \ref{gr}. The wave vector has been scaled with the inverse average distance $\rho^{1/3}$ (the hard sphere $S(q)$ has been left unchanged). The inset shows the neighbour peak in detail. }
\label{sq}
\end{figure}

The scaling observed in Fig. \ref{pressure-scaling}, in contrast with the complex situation observed in Fig. \ref{pressure}, indicates that the relevant length scale for thermodynamic properties is not $\sigma$, but the average particle separation, $\Gamma^{-1/3}=\rho^{-1/3}/\sigma$. This can be observed in the structure of the system, which can be scaled for different states and systems when $\rho^{-1/3}$ is used as the unit of distance. In Fig. \ref{gr} the pair distribution function is presented for the freezing density at every softness. The inset shows the raw data with distance measured in units of $\sigma$, where the maxima are shifted to lower distances for higher $n$, and the main panel presents $g(r)$ with the distance scaled with the mean interparticle distance, where scaling has been achieved for all peaks (in order to keep the first neighbour peak in HS at $r=\sigma$, distances have been scaled not simply by $\Gamma^{-1/3}$, but by $\Gamma_{\text{rel}}^{-1/3}=\left(\Gamma/\Gamma^{HS}\right)^{-1/3}$, where $\Gamma^{HS}$ denotes the hard sphere value of $\Gamma$ at freezing). Differences in the local structure are still observed in the first neighbour peak, caused by the differences in the potential softness, and these are responsible for the differences observed in the pressure (Fig. \ref{pressure-scaling} -- upper panel).

The relevance of the interparticle distance in contrast to $\sigma$ for structural properties, can also be shown using the structure factor, which is more accessible experimentally than the pair distribution function. Fig. \ref{sq} presents the structure, $S(q)$, as a function of the wave vector scaled with the inverse separation distance $\rho^{1/3}$ at freezing for different potential softness. The inset shows the neighbour peak in more detail, where the differences in the local structure should be noticed. As expected from the Hansen-Verlet criterion \cite{Hansen69}, the height of the peak is around 2.85 in all cases, but no difference in the shape of the curves or position of the maximum can be observed beyond the noise level. This shows that all systems in the range of softness studied have similar liquid structure (except locally) at their freezing points, and that the scaling factor is the mean interparticle distance, thus establishing the density that must be used to match the thermodynamic quantities in systems with different softnesses.

We move now to the study of dynamic quantities, particularly the transport coefficients (diffusion coefficient and the shear viscosity), and the autocorrelation functions involved in their calculations. The diffusion coefficient, $D$, can be obtained via the Green-Kubo relation from the velocity autocorrelation function, or from the long time slope of the mean squared displacement\cite{hansen86}, $\langle\delta r^2\rangle$:

\begin{equation}
D\:=\:\int_0^{\infty} C_v(t) dt \: = \:\lim_{t\rightarrow \infty} \frac{\langle \delta r^2(t) \rangle }{6 t}
\end{equation}

\begin{figure}
\psfig{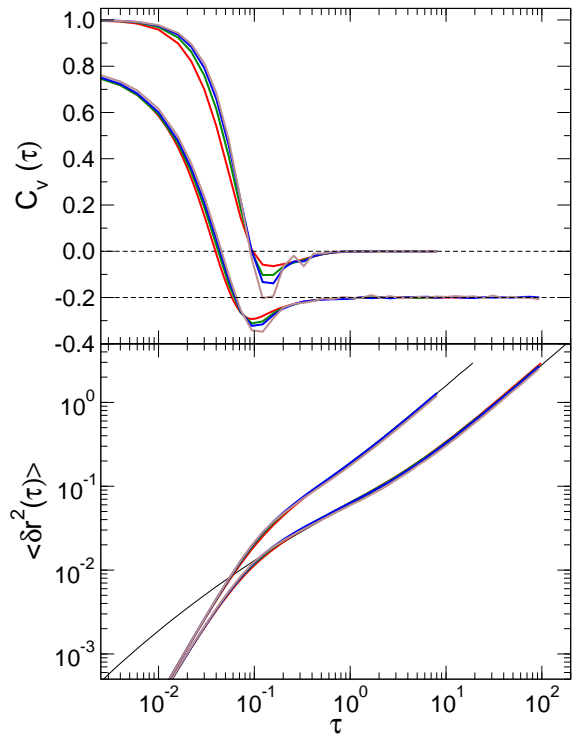}
\caption {Velocity autocorrelation function (upper panel) and mean squared displacement (lower panel) for SS with $n=36$ (red), $n=24$ (green), $n=18$ (blue) and $n=10$ (brown) and HS (only the MSD) (thin black line) at the freezing density (see table I) with ND (upper set of curves in both panels) and BD or dND (lower set of curves). The $C_v(t)$ for dND have been displaced vertically for clarity. BD has been rescaled on the time axis by a factor $\tau_{D}=0.1132$ such that the long time diffusion agrees.}
\label{msd}
\end{figure}

\noindent where $C_v(t)=1/3\; \langle \vec{v}_i(t)\,\cdot\,\vec{v}_i(0)\rangle$ while $\delta \vec{r}=\vec{r}(t)-\vec{r}(0)$, and the brackets indicate ensemble and time-origin averaging. The velocity autocorrelation function is presented in Fig. \ref{msd} (upper panel), for SS with Newtonian dynamics (ND) and damped Newtonian dynamics (dND), at the freezing point for every system. $C_v(t)$ decays faster for dND due to the random forces and friction with the solvent, resulting in lower diffusion coefficient than ND. (The presence of backwards motion indicates that the Brownian forces are not strong enough to provoke the decorrelation of velocities before collisions occur). Because the local environment of the particles is different for the different systems, the $C_v(t)$ functions do not coincide in the decay and at the backscattering minimum, which is deeper for softer potentials. These differences however concur to give equal diffusion coefficients for different particle softness, as shown below in Fig. \ref{diff}.

The mean squared displacement (MSD) is plotted in the lower panel of the figure for SS with ND and dND, and for HS with ND and Brownian dynamics (BD). Due to the Brownian forces (in dND) or kicks (in BD), the MSD grows slower for dND or BD than for ND. (The short time dynamics for HS with BD is controlled by the arbitrary time interval between Brownian kicks, $\tau_B$, and the MSD in this case has been shifted in time to give the same long time self diffusion coefficient). Note that the MSD with the same microscopic dynamics collapse independently of $n$ (except for HS with BD), even though the density is rather different. The velocity autocorrelation functions did not collapse because this function focuses on the microscopic dynamics where the local structure is sampled, although the final diffusion coefficient is identical for all potentials. The scaling in MSD, and the concomitant agreement of diffusion coefficients, indicates again that the relevant quantity for this transport coefficient is the distance to freezing, instead of the density.

\begin{figure}
\psfig{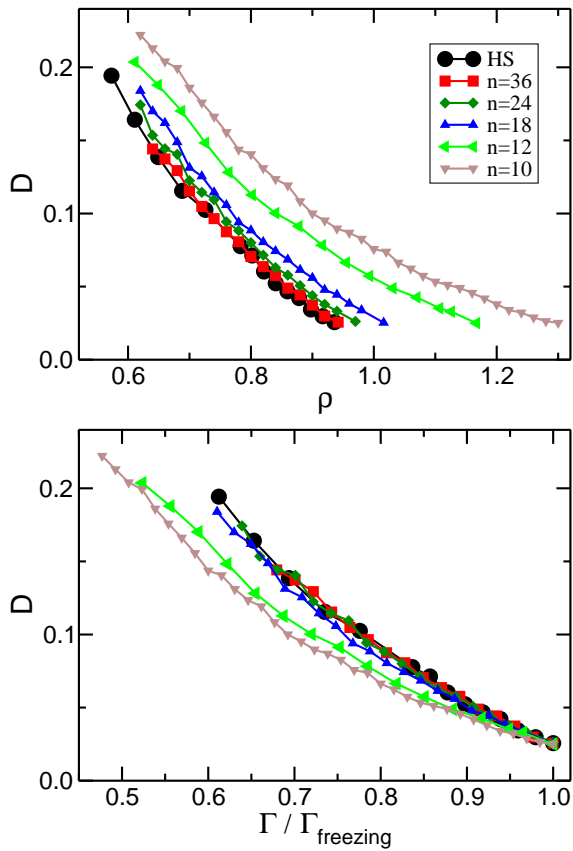}
\caption {Diffusion coefficient for increasing density approaching the freezing transition for soft spheres with different softness and hard spheres, as a function of the density, $\rho$ (upper panel) and the scaled density, $\Gamma/\Gamma_{\rm freezing}$ (lower panel). From top to bottom in the upper panel (bottom to top in the lower one): $n=10$, $n=12$, $n=18$, $n=24$, $n=36$ and hard spheres.}
\label{diff}
\end{figure}

The diffusion coefficient from the long time limit of the MSD with ND for the different potentials is presented in Fig. \ref{diff}, as a function of the density (upper panel), and as a function of $\Gamma/\Gamma_{\rm freezing}$ (lower panel). Whereas the raw data shows differences due to the softness of the potential up to $n=36$ (upper panel), in agreement with previous results \cite{Heyes05}, the value of the diffusion coefficient at the freezing point is very similar in all cases, as expected from the dynamic criterion for freezing proposed by L\"owen et al. \cite{Lowen93} (note, however, that a short-time diffusion coefficient cannot be defined for ND). When the diffusion coefficient is plotted against $\Gamma/\Gamma_{\rm freezing}$, a convincing collapse of the data is found for $n \geq 24$, with small deviations for $n=18$ (lower panel). The same scaling that was used to rationalise the structural and thermodynamic properties is successful in the collapse of the diffusion coefficient. Similar collapse of the diffusion coefficient can be found using the BD or dND as short time dynamics.

Finally, we study the viscosity of the system, which can be calculated from the stress autocorrelation function, $C_{\sigma\sigma}(t)$. The stress tensor is calculated as:

\begin{equation}
\sigma^{\alpha\beta}\:=\:\sum_{i=1}^Nm v_{i\alpha}v_{i\beta}\,+\,\sum_{i<j}^N r_{ij\alpha} F_{ij\beta},
\end{equation}

\begin{figure}
\psfig{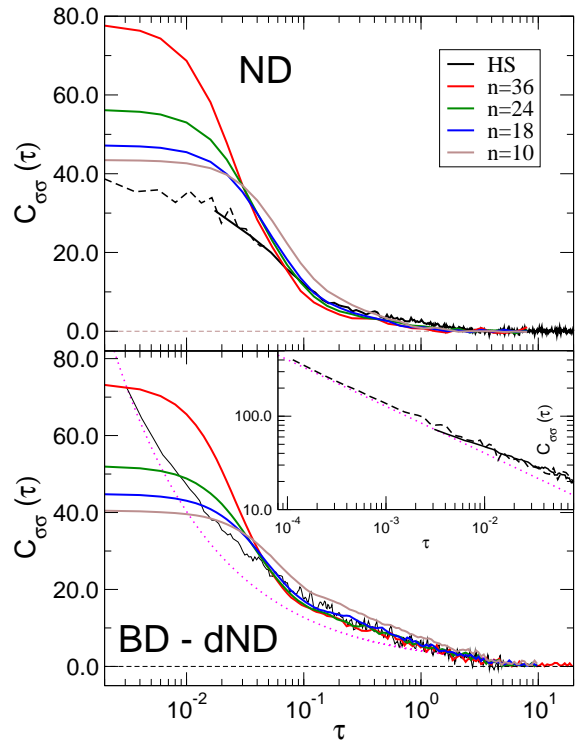}
\caption {Stress correlation function at the freezing density for different potential softness with Newtonian dynamics (upper panel) and Brownian dynamics (lower panel); from top to bottom $n=36$, $n=24$, $n=18$, $n=10$ and hard spheres (black). Dashed black line in the upper panel refers to simulations with $\tau_{\sigma}=0.0017$ to probe shorter correlation times. The inset shows the convergence to $1/\sqrt{t}$ (magenta dotted) for sufficiently short time. To probe these timescales a Brownian time step $\tau_{B}=0.0017$ was used (dashed line, inset).}
\label{csigma}
\end{figure}

\noindent where $v_{i\alpha}$ is the $\alpha$-th component of the velocity of particle $i$, and $F_{ij\beta}$ is the $\beta$-th component of the interaction force between particles $i$ and $j$. 

For HS, however, the interaction potential is not differentiable, and the force cannot be derived from it, although there are collisions between particles. The interaction force can be found using the exchange of momentum between colliding particles\cite{Alder57, Foss00}. To this end one needs to average the change of momentum  $F_{ij\beta}=m\Delta v_{ij\beta}\delta(t-t_{ij})$ during a brief time interval $\tau_{\sigma}$, where the $\delta$-function refers to the instantaneous change of momentum, which leads to
\begin{equation}
\sigma^{\alpha\beta}\:=\:\sum_{i=1}^Nm v_{i\alpha}v_{i\beta}\,+\,\frac{1}{\tau_{\sigma}}\sum_{coll}^{\tau_{\sigma}} m r_{ij\alpha} \Delta v_{ij\beta},
\end{equation}
where the second summation now refers to all collisions within the time interval $\tau_{\sigma}$. Unless otherwise stated $\tau_{\sigma}=0.017$.

The stress correlation function, $C_{\sigma\sigma}(t)=1/3 \sum_{\alpha<\beta} \langle \sigma_{\alpha \beta}(t)
\sigma_{\alpha\beta}(0) \rangle$, is presented in Fig. \ref{csigma} for the hard and soft spheres with different softness at their freezing points, using ND and BD or dND (the brackets indicate average over time origin). In both microscopic dynamics, differences are noticed in the correlation function at short time, but they collapse within the noise level at longer times (compared to the microscopic time scale) for $n \geq 24$. The curve for HS with BD has been scaled in time by the same factor as the MSD (see Fig. \ref{msd}, lower panel), and the collapse with SS is observed for $\tau>10^{-1}$, as in the MSD case.

The short time divergence of the HS correlation function can be observed in the inset of the bottom panel. The coefficient to the $t^{-\frac{1}{2}}$-divergence --- indicated by the magenta dashed line --- has been determined theoretically \cite{Lionberger94,Verberg97}:

\begin{equation}
C_{\sigma\sigma}(t\to 0) =
\frac{18\eta_{0}}{5\tau_{a}}\phi^{2} g(\sigma)\sqrt{\frac{2}{\pi}}\sqrt{\frac{\tau_{a}}{t}}, \label{eq:CssZeroHS}
\end{equation}

\noindent where $\tau_{a}=\frac{\sigma^{2}}{4D_{0}}$ with $D_{0}$ being the short time diffusion coefficient. (These short times were only accessible by reducing the Brownian time scale $\tau_{B}$ by a factor of 10.)

\begin{figure}
\psfig{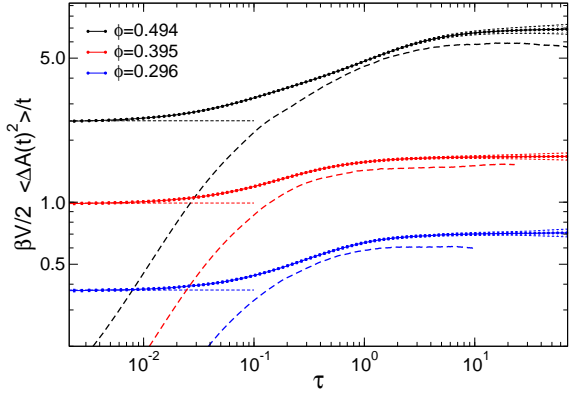}
\caption {Calculation of the viscosity from the squared integrated stress $\Delta A(t)$ at different densities, indicated by color. Solid line with filled symbols show HS, dashed lines indicate soft spheres for $n=36$. Thin dashed lines mark the theoretical result of the $\delta$-contribution  in the stress autocorrelation function with the weight given in eq. (\ref{etaDelta}).}
\label{etaInt} 
\end{figure}

For the case of ND, on the other hand, it has been shown \cite{Dufty02} that the stress correlation function of HS has two contributions: a $\delta$-function with weight $\eta_{\delta}$ at the origin and a well behaved bounded curve for the remaining region. We sample the latter part only.

Similar to the diffusion coefficient, the viscosity can be calculated via a Green-Kubo relation with the stress correlation function, or by the long-time limit of the integrated stress:

\begin{equation}
\eta\:=\:\beta V\int_0^{\infty} dt\,C_{\sigma\sigma}(t) \:=\:\frac{\beta V}{2}
\lim_{t \rightarrow \infty} \frac{1}{t} \langle \Delta A(t)^2 \rangle,
\label{eta}
\end{equation}

\noindent where $\Delta A(t)=\frac{1}{3} \int_0^{t} \sum_{\alpha < \beta} \sigma^{\alpha\beta}(t') dt'$, and the brackets indicate averaging over time origin for this integral, and $V$ is the volume of the system. In HS, the $\delta$-function in $C_{\sigma\sigma}(t)$ mentioned above gives a contribution to the viscosity at $t=0$ equal to $\eta_{\delta}$ with\cite{Branka04}:

\begin{equation}
\eta_{\delta}\:=\:\eta_{0}\frac{192}{25\pi}\left(\frac{P\pi}{6k_{B}T}-\phi\right),
\label{etaDelta}
\end{equation}

\begin{figure}
\psfig{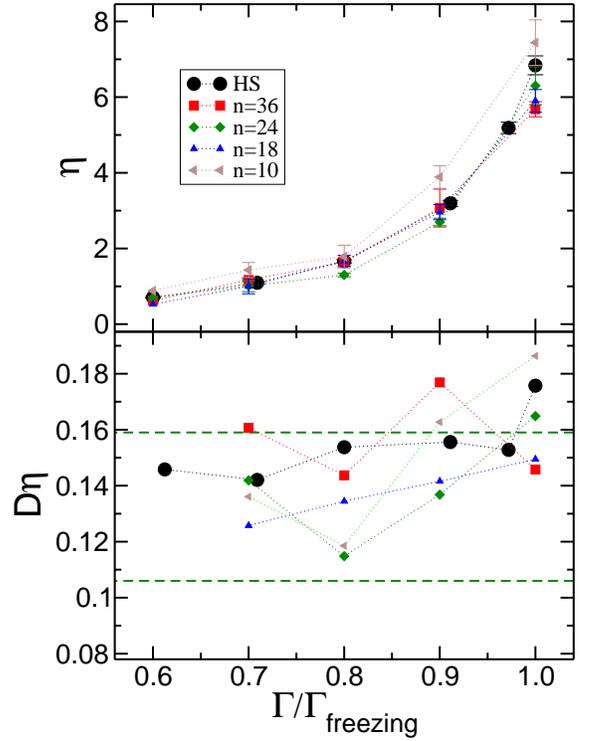}
\caption {Viscosity of hard and soft spheres from the long time slope of the integrated stress, with microscopic ND (upper panel). Different potential softness is studied: black points, hard spheres; red ones, $n=36$; green ones, $n=24$; blue ones, $n=18$; orange ones, $n=10$. The lower panel shows the Stokes-Einstein relation $D\eta$, with the same colour code, and the dashed green thick lines marks the slip value $1/(2\pi)$ (upper line) and stick one $1/(3\pi)$ (lower line).}
\label{viscosity}
\end{figure}

\noindent where $P$ is the pressure and $\eta_0$ is the viscosity from kinetic theory $\eta_{0}=\frac{5}{16\sigma^{2}}\left(\sqrt{\frac{mk_{B}T}{\pi}}\right)$. Numerically, the integration of the correlation function is noisier and therefore, the viscosity is usually calculated with the integral of the stress tensor (second equality in eq. (\ref{eta})), presented in Fig. \ref{etaInt} -- 10000 independent calculations of $\eta$ have been performed and averaged. Note the finite short time limit for HS, which agrees with the theoretical value $2\eta_{\delta}$, in stark contrast to the decay to zero for SS. This window of qualitative difference between SS and HS moves to shorter times upon increasing $n$. The viscosity, long time plateaus in Fig. \ref{etaInt}, are finally presented in Fig. \ref{viscosity} for hard and soft spheres with microscopic ND, as a function of $\Gamma/\Gamma_{\rm freezing}$. The error bars are the standard deviation of the averaging. (Note that Fig. \ref{etaInt} is in logarithmic scale, what enlarges the difference between HS and SS at small values of the viscosity.)

Due to the noise of the data, the collapse is not as impressive as for previous quantities. Deviations are only noted beyond the noise level at high density, close to $\Gamma_{\rm freezing}$, for $n=10$. These results are compatible with the previous ones in showing that the appropriate scaling of the thermodynamic quantities and transport coefficients is obtained when $\Gamma$ is referred to its value at freezing, although definite proof is only obtained for previous quantities. We have included it here, nevertheless, because of its relevance to experiments in colloid science.

The lower panel of the figure shows the product $D \eta$, which would be constant according to the Stokes-Einstein relation. Surprisingly, as this relation derived from continuum fluid mechanics need not apply to the present situation of the motion of a particle surrounded by a fluid of identical particles, within the noise level, this product is indeed constant, and coincides with the theoretical value of the Stokes' law for a sphere with slippery surface $D \eta= (2 \pi \sigma)^{-1}$ for $n=36$.

\section{Conclusions}

We have studied in this work the structural and transport properties
of hard and soft spheres with different softness using simulations
with different microscopic dynamics. The pressure and the transport
coefficients can be rescaled onto the HS values at high density when
studied as a function of the density normalized to the freezing
point for every system for $n\geq 18$, although they typically show
non-monotonous behaviour for increasing density (or volume
fraction), irrespectively of the microscopic dynamics.
Concomintantly, distances must be measured in units of the
interparticle distance, and all systems have the same long range
structure at their freezing points. Differences due to the different
softness are appreciable only in the local environment of the
particles, and at short times in the velocity and stress
autocorrelation functions. The latter is particularly interesting
since its value at zero-time depends on the particle softness and
diverges for hard spheres. The analytical prediction at short times
for the stress autocorrelation function of hard spheres and the
initial $\delta$-contribution to the viscosity have been verified
quantitatively with our simulations. As expected, differences in the
correlation functions are also noticed depending on the microscopic
dynamics used, although the final values of the transport
coefficients can be scaled.

These results show that the key parameter for dense systems of hard
and soft spheres is not the density, but the relative distance to
the freezing point (which itself shows a non-monotonous behaviour
with the softness of the potential). This can be of special interest
for experimentalists, where a small softness cannot be ruled out, in
particular for particles covered with surfactant layers, or in cases
where the determination of the colloid density poses a major
problem. Importantly, and perhaps at first counter intuitively, our
simulations show that the mapping of a fluid of soft spheres onto
the one of hard spheres becomes progressively better at higher
densities and closer to freezing. This supports the idea, going back
to van der Waals, that the local packing of a dense fluid is
dominated by the rapidly varying repulsion which can successfully be
modeled by the hard sphere one \cite{Andersen71}. In these cases,
our results indicate that the relevant quantity to compare different
systems is not the bare density, but its ratio to the freezing
density. In passing, we also note that this scaling is much simpler
and more effective than other attempts to map the structure and
dynamics of soft spheres onto those of hard spheres, based on
effective hard-core diameters (as far as the particle is not
extremely soft, i.e. $n \geq 18$).

\begin{center}
{\sc Acknowledgments}
\end{center}

We acknowledge helpful discussions with C. di Michele, A. Scala, Th. Voigtmann, and F. Weysser. Financial support for A.M.P. was provided by the Spanish Ministerio de Educaci\'on y Ciencia, project MAT 2006-13646-C03-02. J.B.C. and E.L. were partially funded by the Acciones Integradas / DAAD program, Ref. HA2004-0022.

\clearpage



\end{document}